# Rotational-state-selected Carbon Astrochemistry

Jutta Toscano*

*Abstract:* The addition of individual quanta of rotational excitation to a molecule has been shown to markedly change its reactivity by significantly modifying the intermolecular interactions. So far, it has only been possible to observe these rotational effects in a very limited number of systems due to lack of rotational selectivity in chemical reaction experiments. The recent development of rotationally controlled molecular beams now makes such investigations possible for a wide range of systems. This is particularly crucial in order to understand the chemistry occurring in the interstellar medium, such as exploring the formation of carbon-based astrochemical molecules and the emergence of molecular complexity in interstellar space from the reaction of small atomic and molecular fragments.

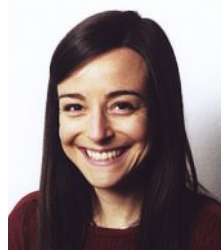

*Jutta Toscano* was born in Siena (Italy) in 1990 and studied chemistry at University College London (UK), where she graduated with an MSci degree in 2014. She received a DPhil in Physical and Theoretical Chemistry in 2018 from the University of Oxford (UK) where she completed her doctoral studies under the supervision of Professors Tim Softley and Brianna Heazlewood. She was awarded an EPSRC Doctoral Prize Fellowship in 2018 and a Lindemann Postdoctoral Fellowship in 2019 when she moved to JILA, University of Colorado Boulder (USA), to work in the group of Prof. Jun Ye. In 2021, she joined the group of Prof. Stefan Willitsch at the University of Basel (Switzerland), and then in 2023 she started her independent group as an SNSF Ambizione Fellow.

## 1. Introduction

Chemical reactions generally occur between molecules in a variety of quantum states, with different kinetic energies and orientations and can be compared to an orchestra performing a symphony, with all instruments playing their individual tunes simultaneously. To gain a deeper understanding of the symphony, we wish to listen to the individual instruments playing unaccompanied, in order to appreciate each of their tunes in detail. In the chemical world, this is equivalent to selectively observing the reactivity of molecules in a specific electronic, vibrational and rotational state. However, studying a perfectly controlled chemical reaction is not trivial since the theoretical simplification of the problem translates into experimental complexity. Preparing the reagents such that each molecule is present in a specific quantum state and possesses a specific amount of kinetic energy requires the employment of an array of techniques that belong to the intersection between chemistry and physics.[1–5] In turn, the impact of investigating perfectly controlled reactions spans across diverse fields from fundamental reaction dynamics, through molecular physics to astrochemistry and astrophysics, making this a highly interdisciplinary endeavour.

### 1.1 *Carbon Astrochemistry*

Space is a highly complex, diverse and constantly evolving environment which we mainly study through remote observation in the ongoing quest to understand it. The molecules present in space are exceptionally important to this end since they permit, through spectroscopy, a glimpse of the changing conditions of various regions of space such as the temperature and density of diffuse and dense clouds in the interstellar medium.[6–9] Indeed, molecules have been detected across a wide variety of regions of space and the study of their formation, reactivity and destruction in harsh environments that are so different from Earth is a fascinating pursuit in its own right. In contrast to the chemistry on Earth, reactivity in space is primarily driven by kinetics instead of thermodynamics due to the intricate interplay between radiative and collisional processes, leading to a rich organic inventory of exotic unsaturated molecules and unusual isomeric abundances.[6] Additionally, highly reactive and excited species are abundant due to the relative ease with which they are formed, through interaction with stellar UV radiation and cosmic rays, and the reduced probability of being destroyed, given the very low gas densities which lead to infrequent collisions. Knowledge of the rates of reactions that occur in space is necessary in order to accurately model its chemistry. For instance, reliable models of the interstellar medium, in which the predicted abundances of all species agree with the experimentally observed abundances, can be used to explain and predict the evolution of different molecular clouds, to simulate stellar lifecycles (Fig. 1) or even to investigate the possible development of prebiotic molecules from simple organic species, which could explain the origin of life. Of particular relevance to the latter is the study of carbon chemistry and how molecular complexity emerges from the reaction of small atomic and molecular fragments that are present in space.

Most of the carbon in space is locked up in very stable species, with the two main reservoirs being CO and PAH (Polycyclic Aromatic Hydrocarbon) molecules.[6] Consequently, the abundant organic inventory of the solar system mostly derives from two independent routes: one that builds complex species starting from CO in interstellar clouds and the other that breaks down complex PAHs from stellar regions into smaller species. Here we focus on the former bottom-up route, where gas-phase and grain-surface chemistry in diffuse and dense interstellar clouds drive the formation of hundreds of increasingly complex carbon-bearing molecules from small, single-carbon species. The gas-phase chemistry of these clouds is generally dominated by ion–mole-

*Correspondence:* Dr. J. Toscano, E-mail: jutta.toscano@unibas.ch
Department of Chemistry, University of Basel, CH-4056 Basel

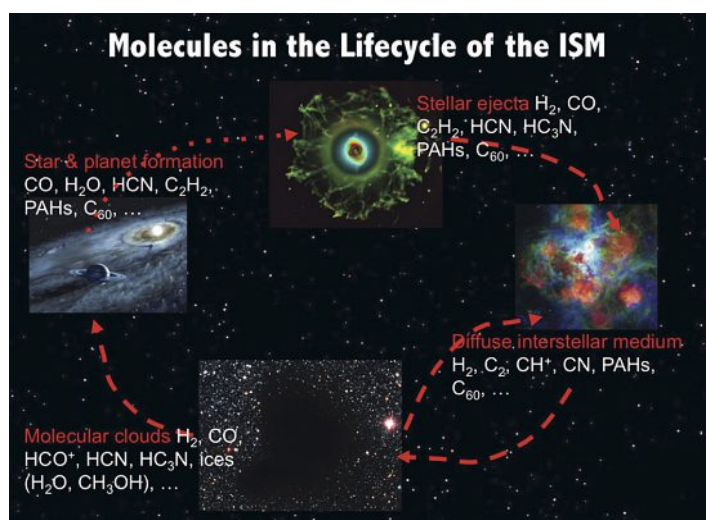

Fig. 1. In the interstellar medium (ISM), the various stages of the stellar lifecycle are punctuated by the presence of different molecules which facilitate their characterisation as well as our understanding of their distinct chemistry. Reproduced with permission from the APS.[6]

cule reactions and their composition is simulated using kinetic models consisting of large networks of reactions (~5000 reactions between ~450 species). These models rely heavily on databases such as UDfA[10,11] and KIDA[12,13] to supply experimentally measured or theoretically calculated reaction rate coefficients and product branching ratios for the multitude of chemical processes involved. Unfortunately, less than 20% of the reactions included in these databases have actually been studied in the laboratory and this causes order-of-magnitude discrepancies between simulated and observed abundances.[6] Incorrect reaction rates or branching ratios for one reaction can have dramatic ripple effects in the estimated abundances of hundreds of other species and, therefore, the lack of laboratory measurements of these quantities severely hampers detailed modelling efforts.

Particularly urgent is the need for quantum-state-resolved studies since kinetics tends to dominate over thermodynamics in space, as discussed above, and even rotational effects can have drastic kinetic consequences at low temperatures.[14,15] Indeed, despite the small energy difference between rotational levels, the addition of a single quantum of rotational excitation has already been demonstrated to change (i) Penning ionisation rate,[16] (ii) reaction rate[17] and (iii) product branching ratio[18] of different reactions, as detailed in Section 1.2. The distinct chemical behaviour of molecules in different quantum states clearly demonstrates the need to treat them as distinct chemical species, particularly when modelling collision-scarce environments that are not at thermal equilibrium, such as interstellar clouds. The lack of state selectivity in the vast majority of the reaction studies carried out so far is due to the experimental challenges associated with selectively generating and maintaining the reagents in a specific quantum state or, alternatively, purifying an ensemble of molecules such that the ones partaking in the reaction are in a single quantum state. However, the remarkable advances of the past few decades in the state-selective manipulation of an increasing variety of atomic and molecular species with diverse properties renders such experiments possible now.

### 1.2 Rotational-state-selected Chemistry

Experimental techniques such as supersonic expansion, electric or magnetic deceleration or deflection, trapping of ionic and neutral species, laser cooling and optical pumping are frequently combined in creative new ways to advance our understanding of fundamental reaction dynamics.[1–5] Exceptional control over the collision energy and quantum state can be achieved in merged and crossed beam experiments, and this has permitted the detailed probing of long-range intermolecular forces and the observation of quantum effects such as tunnelling and scattering resonances. Trapping of one or both reacting partners has extended the observable interaction times to the seconds and minutes time-scales, drastically enhancing the sensitivity and resolution of reaction rate measurements. Lasers are commonly used to selectively populate, as well as selectively probe, specific quantum states. Despite the significant progress made thus far in extending the scope of these techniques (and combination of techniques) to manipulate an increasing variety of systems with diverse properties, very few experiments have been able to directly probe the effect of rotational excitation on chemical reactivity to date.[16–21] This is partly due to the ease with which collisions or absorption of black-body radiation can prompt a change in the rotational state of a molecule, given the small energy differences between rotational levels, leading to undesirable scrambling of states and loss of selectivity. Another factor that plays into the scarcity of rotational-state-resolved experiments is the challenge of purifying the sample such that all molecules partaking in the reaction are in a single rotational state.

The rotationally selective experiments that have been performed so far show fascinating effects and point towards an abundant body of knowledge yet to be uncovered. For instance, the rotationally excited spin isomer ortho-$H_2$ (rotational quantum number, $j = 1$) has been found to ionise faster than the ground state para-$H_2$ ($j = 0$) in Penning ionisation reactions with metastable helium within a merged beam experiment.[16,22] The addition of one quantum of rotational excitation modifies the long-range intermolecular interaction potential, from being isotropic to an-isotropic, thereby strengthening the long-range attractive forces. Extracting such remarkable levels of detail from the collision process is only possible with very precise control over the reaction parameters; specifically, the quantum states and collision energy of the colliding species. In this case, a pure sample of para-$H_2$ was obtained by liquifying normal-$H_2$ ($j = 0,1$) in the presence of a catalyst, with the contribution from ortho-$H_2$ inferred from the difference in reactivity between para-$H_2$ and normal-$H_2$. A more general and widely applicable method to isolate the contribution of different rotational states has been developed:[23,24] an electrostatic deflector that spatially separates different rotational states of a polar molecule by exploiting their different interactions with an applied inhomogeneous electric field. The force the molecules experience inside the deflector, and therefore the amount by which they are deflected in space, is quantum-state dependent since different rotational states have different effective dipole moments, as discussed in more detail in Section 2.1. This technique has enabled the investigation of the reactivity of ortho- and para-water towards diazenylium ions:[17] $H_2O + N_2H^+ \rightarrow N_2 + H_3O^+$. The ground state spin isomer para-$H_2O$ ($j = 0$) was shown to react faster than the rotationally excited ortho-$H_2O$ ($j = 1$) due to the smaller degree of rotational averaging of the ion-dipole long-range intermolecular interaction. Recently, the branching ratio of the reaction between OCS and metastable Ne has also been observed to depend on the rotational state of the molecule, with dissociative ionisation occurring faster for OCS ($j = 0$) compared to OCS ($j = 1$).[18] Overall, besides being theoretically predicted to significantly decrease reactivity at low temperatures,[25] the addition of a single quantum of rotational excitation has so far been experimentally demonstrated to change (i) ionisation rate, (ii) reaction rate and (iii) product branching ratio of different systems. This evidence challenges the implicit assumption, that interstellar models currently make, that molecules in different rotational states have identical reactivity, and it simultaneously stimulates further research efforts into quantum-state-selected reaction dynamics.

## 2. Experimental Methods

The high degree of control necessary to investigate rotational-state-resolved ion–molecule reactions requires an apparatus that combines the ability to separate molecules in different rotational states while studying their reactivity towards ionic species. This is achieved by combining an electrostatic deflector with a linear-quadrupole radio-frequency ion trap. The deflector separates the neutral molecules such that only those in a selected rotational state collide with the ionic reactant held in the trap (Fig. 2). The force the molecules experience inside the deflector, and therefore the amount by which they are deflected in space, is quantum-state dependent since different rotational states have different effective dipole moments. The progress of the reaction is monitored using a camera whilst the identity and amounts of ionic products are established by mass spectrometry as a function of time, enabling the extraction of reaction rate coefficients and branching ratios for all product channels (Fig. 3), as explained in more detail below.

The electric field manipulation and ion trapping require collision-less high- and ultra-high-vacuum environments, respectively. For this reason, the neutral reactants are introduced into the experimental chamber in the form of a gas-phase molecular beam. The neutral reactant is seeded into a noble gas and supersonically expanded through a nozzle. The adiabatic cooling that takes place efficiently quenches the rotational energy of the molecules, collapsing the population to the lowest rotational states, as well as drastically decreasing their translational velocity spread down to a few Kelvin. This comes at the expense of an increased kinetic energy of the beam in the laboratory frame, which is typical of supersonic beams. The molecular beam is collimated by two skimmers before passing through the electrostatic deflector, where an inhomogeneous electric field spatially separates the polar molecules based on their rotational quantum state.

### 2.1 *Electrostatic Deflection*

The Stark effect lifts the degeneracy of a molecule's quantum state based on the relative orientation of its dipole moment with respect to the direction of the applied field. Molecules with the dipole moment parallel to the electric field experience a decrease in energy that is proportional to the magnitude of the applied field. These are referred to as high-field-seeking since they are attracted towards high electric field regions where their energy is lower. On the contrary, molecules with the dipole moment antiparallel to the electric field are known as low-field-seeking, since their energy increases with increasing field strength and, consequently, they are attracted by regions of lower field strength. The electrostatic deflector featured in this experiment addresses high-field-seeking molecules and works by attracting them towards a high electric field region. This serves the double purpose of purifying the molecular sample – by attracting the desired particles away from the beam axis where the rest of the molecules and seed gas are – and broadly sorting the molecules in order of increasing rotational quantum number.[24]

The inhomogeneous electric field generated inside the deflector interacts with the dipole moment of the polar molecules in a rotational-state-specific manner. The rotational motion of the molecules modulates their effective dipole moment and, therefore, the strength of their interaction with the applied field depends on their rotational quantum number, $j$. Molecules in the lowest rotational state ($j = 0$) typically have the largest effective dipole moment, the strongest Stark interaction with the applied field and, hence, they are deflected the most.[24] The amount of deflection and state separation achievable depends on the dipole-moment-to-mass ratio of the molecule – alongside the molecule-specific rotational structure detail. In general, lighter molecules with larger dipole moments experience a stronger deflecting force which leads to better state separation. New potential candidates for deflection need to be assessed individually and their deflection carefully simulated to determine whether sufficient separation between molecules in different rotational states is feasible. In practice, only the lowest two or three rotational states are usually deflected enough to be separated from the main molecular beam.

In the case of water (dipole moment, $\mu = 1.855$ D),[26] the two lowest rotational quantum states also correspond to two distinct spin isomers which differ by their nuclear spin quantum number, $I$, where the relative orientation of the hydrogen nuclear spins is antiparallel for $I = 0$ (*para*) and parallel for $I = 1$ (*ortho*). According to the Pauli principle, which requires the total molecular wavefunction to be antisymmetric with respect to a permutation of any two electrons, the *para* (antisymmetric) spin wavefunctions can only exist in even (symmetric) rotational levels whereas the *ortho* (symmetric) spin wavefunction can only exist in odd (antisym-towards the ionic target. Inside the ion trap, calcium ions are held in place by a combination of radio frequency and DC electric fields, as well as being laser cooled to reach temperatures as low as a few *milli*Kelvin. Below a critical temperature, the trapped ions form so-called 'Coulomb crystals': structures where the gas-phase ions, although still mobile, tend to occupy specific positions in space and resemble the structure of a solid crystal.[27,28] The continuous laser cooling of the $Ca^+$ ions induces a constant fluorescence which allows the direct observation of the Coulomb crystal using a camera. The balance between the confining forces of the trapping fields and the Coulombic repulsion between the positively charged, low-kinetic-energy ions results in neighbouring ions within the crystal being separated by 10–20 μm, with densities in the order of $10^8$ cm$^{-3}$ compared to typical densities of ≈$10^{22}$ cm$^{-3}$ in solid crystals. At the lowest temperatures, their fluorescence can even reveal the average positions of the individual $Ca^+$ ions within the crystal, as shown in the left panel of Fig. 3(a). Other ions can be added to the trap and their excess kinetic energy is removed *via* elastic collisions with the laser-cooled $Ca^+$ ions, which effectively act as a heat sink. This process is referred to as 'sympathetic cooling' and it allows the co-trapped atomic[29] or molecular[30] ions to be incorporated into the Coulomb crystal. Depending on whether the non-fluorescing ions are heavier or lighter than calcium, they either sit around or inside the $Ca^+$ shell, appearing as dark areas due to the lack of fluorescence from the displaced $Ca^+$ ions (Fig. 3(a), right panel). After they are formed (in the timescale of a few minutes) mixed-species Coulomb crystals are stable for extended periods of time (up to hours), offering long interaction times with the incoming neutral molecules and enabling even the examination of reactions that are too slow to be studied in beam or flow-based experiments.[1–4,28]

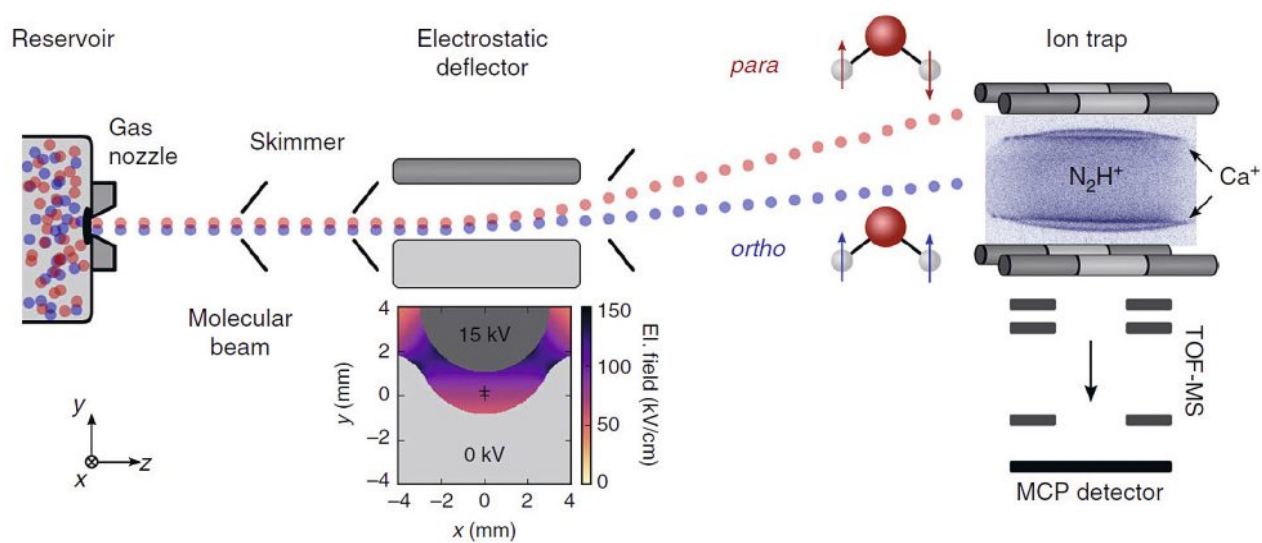

Fig. 2. Schematic of the experimental setup. A pulsed molecular beam of water molecules seeded in argon emanates from a room-temperature reservoir through a pulsed gas nozzle and passes an electrostatic deflector. The inhomogeneous electric field inside the deflector (shown in the inset below) spatially separates para- and ortho-water molecules due to their different effective dipole moments. After the deflector, the beam is directed at an ion trap containing a Coulomb crystal of Ca$^+$ and sympathetically cooled N$_2$H$^+$ reactant ions (inset image). The products and kinetics of reactive collisions between N$_2$H$^+$ and H$_2$O are probed using a time-of-flight mass spectrometer (TOF-MS). Reproduced with permission from Kilaj *et al*.[17]

### 2.3 Reaction Rates and Branching Ratios

The reaction between the ions embedded within the $Ca^+$ Coulomb crystal and the rotationally selected neutral molecules occurs inside the ion trap and can be visually monitored as a function of time from the change in the shape of the crystal, as shown in Fig. 3 for a different reaction system. As the reaction proceeds and the ionic reactants are depleted, the newly formed ionic products get incorporated into the Coulomb crystal. The ionic content of the trap can be probed at any point in time during the reaction by ejecting the Coulomb crystal into a time-of-flight mass spectrometer to reveal the identity of all ions present. The destructive nature of this detection technique requires a new crystal to be formed each time the previous one is probed. Ion numbers can also be determined non-destructively by using molecular dynamic simulations to match the experimentally observed fluorescence pattern, which is mainly suitable for the study of simple reaction systems with limited product channels. The formation of individual ionic products is recorded as a function of time by performing this crystal ejection at various times after the reaction is initiated. From this data, the reaction rates and branching ratios for all product channels can be extracted, as shown in Fig. 3(c), giving a full picture of the reaction process.

### 3. Conclusions

The aim of this project is to experimentally measure reaction rate coefficients and product branching ratios as a function of the rotational excitation of the neutral reactant in ion–molecule reactions involving carbocations as well as rotational-state-selected carbon-containing neutral species. The ability to compare the reactivity of a molecule before and after the addition of a single quantum of rotational excitation is truly exceptional at this moment in time, and it can provide novel insights into the fundamental reaction dynamics of astrochemical processes. Chemical reactivity has been demonstrated to depend on the rotational excitation of the reactants but very few rotational-state-resolved re-metric) rotational levels. Given that the conversion between these isomers is symmetry-forbidden for isolated water molecules, *para*-$H_2O$ ($I = 0, j = 0$) and *ortho*-$H_2O$ ($I = 1, j = 1$) coexist without interconverting in a cold molecular beam, and they can be spatially separated to investigate their distinct reactivity.[17]

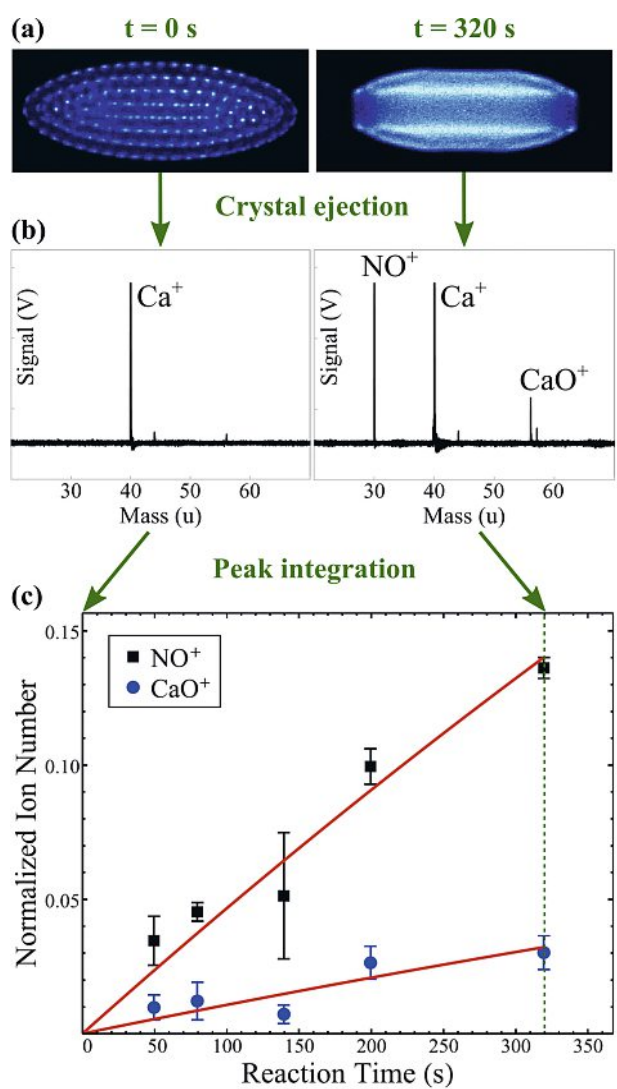

Fig. 3. Schematic representation of the data collection and analysis; here, the $Ca^+$ + NO reaction system is used as an example to illustrate the methodology. The two competing reaction pathways yield $NO^+$ (charge transfer) and $CaO^+$ (O addition) products. (a) Experimental images of the fluorescing calcium ions within a Coulomb crystal are collected as a function of time using a CCD camera. The reaction begins with a crystal that contains only $Ca^+$ ions (left panel); as NO is introduced to the reaction chamber, the accumulation of non-fluorescing, sympathetically cooled ionic products can be inferred from the flattening of the crystal and the appearance of a dark core in its centre (right panel). (b) At various reaction times, all ions within the Coulomb crystal are ejected into a time-of-flight mass spectrometer. The area under each peak in the resulting mass spectrum is proportional to the number of ions of each species at the time of ejection. (c) Fitting the number of product ions as a function of reaction time enables the rate coefficients and product branching ratios to be calculated. Reproduced with permission from Toscano *et al*.[1]

### 2.2 Ion Trapping and Sympathetic Cooling

The purified and deflected beam, with a typical density of $10^8$–$10^9$ cm$^{-3}$, is skimmed once more before reaching the ion trap, where the reaction takes place. Careful tilting of the deflector allows a different rotational state (or spin isomer) to be directed actions have been investigated to date. This lack of experimental data is particularly detrimental for gas-phase kinetic models of the interstellar medium, which are essential to remotely probe the conditions of various regions of space and to understand their chemistry. Combining methodologies for the control and manipulation of ions and neutral molecules, it is now possible to investigate the rotational dependence of crucial astrochemical ion–molecule reactions that lead to carbon–oxygen and carbon–carbon bond formation. This can reveal how molecular complexity in interstellar space emerges from the reaction of small atomic and molecular fragments, improving our ability to accurately model various regions of the interstellar medium, as well as bringing us a step closer to fully controlling the outcome of chemical reactions using electromagnetic fields and laser light.

### *Acknowledgements*

I am grateful to all previous and current mentors, colleagues and collaborators for a decade worth of enriching and stimulating scientific discussions. In particular, I would like to thank Prof. Stefan Willitsch for hosting this Ambizione project within his laboratory.

Received: November 30, 2023

[1] J. Toscano, H. J. Lewandowski, B. R. Heazlewood, *Phys. Chem. Chem. Phys.* **2020**, *22*, 9180, https://doi.org/10.1039/D0CP00931H.
[2] 'Cold Chemistry: Molecular Scattering and Reactivity Near Absolute Zero', Eds. O. Dulieu, A. Osterwalder, The Royal Society of Chemistry, **2017**, https://doi.org/10.1039/9781782626800.
[3] T. P. Softley, *Proc. R. Soc. A.* **2023**, *479*, 20220806 https://doi.org/10.1098/rspa.2022.0806.
[4] 'Chemistry with Controlled Ions, in Advances in Chemical Physics', Ed. S. Willitsch, Ch. 5, pp. 307–340, John Wiley & Sons, Ltd, **2017**, https://doi.org/10.1002/9781119324560.ch5.
[5] B. K. Stuhl, M.T. Hummon, J. Ye, *Annu. Rev. Phys. Chem.* **2014**, *65*, 501, https://doi.org/10.1146/annurev-physchem-040513-103744.
[6] A. G. G. M. Tielens, *Rev. Mod. Phys.* **2013**, *85*, 1021, https://doi.org/10.1103/RevModPhys.85.1021.
[7] W. Klemperer, *Annu. Rev. Phys. Chem.* **2011**, *62*, 173, https://doi.org/10.1146/annurev-physchem-032210-103332.
[8] E. Herbst, E. F. van Dishoeck, *Annu. Rev. Astron. Astrophys.* **2009**, *47*, 427, https://doi.org/10.1146/annurev-astro-082708-101654.
[9] P. Ehrenfreund, S. B. Charnley, *Annu. Rev. Astron. Astrophys.* **2000**, *38*, 427, https://doi.org/10.1146/annurev.astro.38.1.427.
[10] UDfA (The UMIST Database for Astrochemistry), maintained by T. J. Millar and co-workers: http://www.udfa.net/.
[11] D. McElroy, C. Walsh, A. J. Markwick, M. A. Cordiner, K. Smith, T. J. Millar, *A&A* **2013**, *550*, A36, https://doi.org/10.1051/0004-6361/201220465.
[12] KIDA (Kinetic Database for Astrochemistry), maintained by V. Wakelam and co-workers: https://kida.astrochem-tools.org/.
[13] V. Wakelam, E. Herbst, J.-C. Loison, I. W. M. Smith, V. Chandrasekaran, B. Pavone, N. G. Adams, M.-C. Bacchus-Montabonel, A. Bergeat, K. Béroff, V. M. Bierbaum, M. Chabot, A. Dalgarno, E. F. van Dishoeck, A. Faure, W. D. Geppert, D. Gerlich, D. Galli, E. Hébrard, F. Hersant, K. M. Hickson, P. Honvault, S. J. Klippenstein, S. L. Picard, G. Nyman, P. Pernot, S. Schlemmer, F. Selsis, I. R. Sims, D. Talbi, J. Tennyson, J. Troe, R. Wester, and L. Wiesenfeld, *Astrophys. J. Suppl. Ser.* **2012**, *199*, 21, https://doi.org/10.1088/0067-0049/199/1/21.
[14] M. T. Bell, A. D. Gingell, J. M. Oldham, T. P. Softley, S. Willitsch, *Faraday Discuss.* **2009**, *142*, 73, https://doi.org/10.1039/B818733A.
[15] A. Faure, P. Hily-Blant, C. Rist, G. Pineau des Forêts, A. Matthews, D. R. Flower, *MNRAS* **2019**, *487*, 3392, https://doi.org/10.1093/mnras/stz1531.
[16] Y. Shagam, A. Klein, W. Skomorowski, R. Yun, V. Averbukh, C. P. Koch, E. Narevicius, *Nat. Chem.* **2015**, *7*, 921, https://doi.org/10.1038/nchem.2359.
[17] A. Kilaj, H. Gao, D. Rösch, U. Rivero, J. Küpper, S. Willitsch, *Nat. Commun.* **2018**, *9*, 2096, https://doi.org/10.1038/s41467-018-04483-3.
[18] L. Ploenes, P. Straňák, H. Gao, J. Küpper, S. Willitsch, *Mol. Phys.* **2021**, *119*, https://doi.org/10.1080/00268976.2021.1965234.
[19] D. Hauser, S. Lee, F. Carelli, S. Spieler, O. Lakhmanskaya, E. S. Endres, S. S. Kumar, F. Gianturco, R. Wester, *Nat. Phys.* **2015**, *11*, 467, https://doi.org/10.1038/nphys3326.
[20] W. Perreault, N. Mukherjee, R. N. Zare, *Science* **2017**, *358*, 356, https://doi.org/10.1126/science.aao3116.
[21] R. Hahn, D. Schlander, V. Zhelyazkova, F. Merkt, *arXiv:2311.17796*, **2023**, https://doi.org/10.48550/arXiv.2311.17796.
[22] A. Klein, Y. Shagam, W. Skomorowski, P. S. Żuchowski, M. Pawlak, L. M. C. Janssen, N. Moiseyev, S. Y. T. van de Meerakker, A. van der Avoird, C. P. Koch, E. Narevicius, *Nat. Phys.* **2017**, *13*, 35, https://doi.org/10.1038/nphys3904.
[23] L. Holmegaard, J. H. Nielsen, I. Nevo, H. Stapelfeldt, F. Filsinger, J. Küpper, G. Meijer, *Phys. Rev. Lett.* **2009**, *102*, 023001, https://doi.org/10.1103/PhysRevLett.102.023001.
[24] Y.-P. Chang, D. A. Horke, S. Trippel, J. Küpper, *Int. Rev. Phys. Chem.* **2015**, *34*, 557, https://doi.org/10.1080/0144235X.2015.1077838.
[25] D. C. Clary, *Annu. Rev. Phys. Chem.* **1990**, *41*, 61, https://doi.org/10.1146/annurev.pc.41.100190.000425.
[26] NIST Computational Chemistry Comparison and Benchmark DataBase, List of experimental dipoles, https://cccbdb.nist.gov/diplistx.asp; S.L. Shostak, W.L. Ebenstein, J. S. Muenter, *J. Chem. Phys.* **1991**, *94*, 5875, https://doi.org/10.1063/1.460471.
[27] S. Willitsch, *Int. Rev. Phys. Chem.* **2012**, *31*, 175, https://doi.org/10.1080/0144235X.2012.667221.
[28] 'Chemistry Using Coulomb Crystals, in Emerging Trends in Chemical Applications of Lasers', Eds. B. R. Heazlewood, H. J. Lewandowski, Ch. 17, pp. 389–410, ACS **2021**, https://doi.org/10.1021/bk-2021-1398.ch017.
[29] D. J. Larson, J. C. Bergquist, J. J. Bollinger, W. M. Itano, D. J. Wineland, *Phys. Rev. Lett.* **1986**, *57*, 70, https://doi.org/10.1103/PhysRevLett.57.70.
[30] K. Mølhave, M. Drewsen, *Phys. Rev. A.* **2000**, *62*, 011401, https://doi.org/10.1103/PhysRevA.62.011401.